\documentclass[]{aa}
\usepackage{psfig}
\def\apj{ApJ }

\def\aj{AJ }

\def\gsim{\lower.4ex\hbox{$\;\buildrel >\over{\scriptstyle\sim}\;$}}
\def\lsim{\lower.4ex\hbox{$\;\buildrel <\over{\scriptstyle\sim}\;$}}

\begin{document}
%
%
\thesaurus{11.08.1;
           11.09.4;
           11.17.1;
           11.17.4 Q1026-0045A,B
           }
\title{
HST observations of the QSO pair Q1026--0045A,B
\thanks{Based on observations obtained with the NASA/ESA {\sl Hubble Space
Telescope} by the Space Telescope Science Institute, which is operated
by AURA, Inc., under NASA contract NAS 5-26555}}
\author{Patrick Petitjean\inst{1,2} \and Jean Surdej$^{3,7}$ 
\and Alain Smette$^{4}$ \and Peter Shaver$^{5}$ \and Jan M\"ucket$^{6}$
\and Marc Remy$^{3}$}
\institute{$^1$Institut d'Astrophysique de Paris - CNRS, 98bis Boulevard 
Arago, F-75014 Paris, France\\
$^2$UA CNRS 173- DAEC, Observatoire de Paris-Meudon, F-92195 Meudon
Principal Cedex, France \\
$^3$ Institut d'Astrophysique, Universit\'e de Li\`ege, Avenue de
Cointe 5, B--4000 Li\`ege, Belgium \\
$^4$ NASA-Goddard Space Flight Center, Code 681, Greenbelt, MD 20771, USA\\
$^5$ European Southern Observatory, Karl$-$Schwarzschild$-$Str. 2, 
D-85748 Garching bei M\"unchen, Germany \\
$^6$Astrophysikalisches Institut Potsdam, An der
Sternwarte 16, D-14482  Potsdam, Germany\\
$^7$ Directeur de Recherches du FNRS, Belgium}
\date{ }
\offprints{Patrick Petitjean ({\sl petitjean@iap.fr})}
\maketitle
\markboth{}{}
\begin{abstract}
The spatial distribution of the Ly$\alpha$ forest is studied 
using new HST data for the quasar pair Q~1026--0045 A and B
at $z_{\rm em}$~=~1.438 and 1.520 respectively. The angular separation
is 36~arcsec and corresponds to transverse linear separations 
between lines of sight of $\sim$300$h^{-1}_{50}$~kpc ($q_{\rm o}$~=~0.5) over
the redshift range  0.833~$<$~$z$~$<$~1.438. 
From the observed numbers of 
coincident and anti-coincident Ly$\alpha$ absorption lines,
we conclude that, at this redshift, the Ly$\alpha$ structures have typical 
dimensions of $\sim$500$h^{-1}_{50}$~kpc, larger than the mean separation 
of the two lines of sight. The velocity difference, $\Delta V$,
between coincident lines
is surprisingly small (4 and 8 pairs with $\Delta V$~$<$~50 and
200~km~s$^{-1}$ respectively). \\
Metal line systems are present at $z_{\rm abs}$~=~1.2651 and 1.2969
in A, $z_{\rm abs}$~=~0.6320, 0.7090, 1.2651 and 1.4844 in B.
In addition we tentatively identify a weak Mg~{\sc ii} system
at $z_{\rm abs}$~=~0.11 in B.
It is remarkable that the $z_{\rm abs}$~=~1.2651 system is common to
both lines of sight. The system at $z_{\rm abs}$~=~1.4844 
has strong O~{\sc vi} absorption.\\
There is a metal-poor associated system at $z_{\rm abs}$~=~1.4420 
along the line of sight to
A with complex velocity profile. We detect a strong Ly$\alpha$
absorption along the line of sight to B redshifted by only 300~km~s$^{-1}$
relatively to the associated system.
It is tempting to interpret this as the presence of a disk of radius 
larger than 300$h^{-1}_{50}$~kpc surrounding quasar A.

\end{abstract}


\begin{figure}
\centerline{\hbox{\psfig{figure=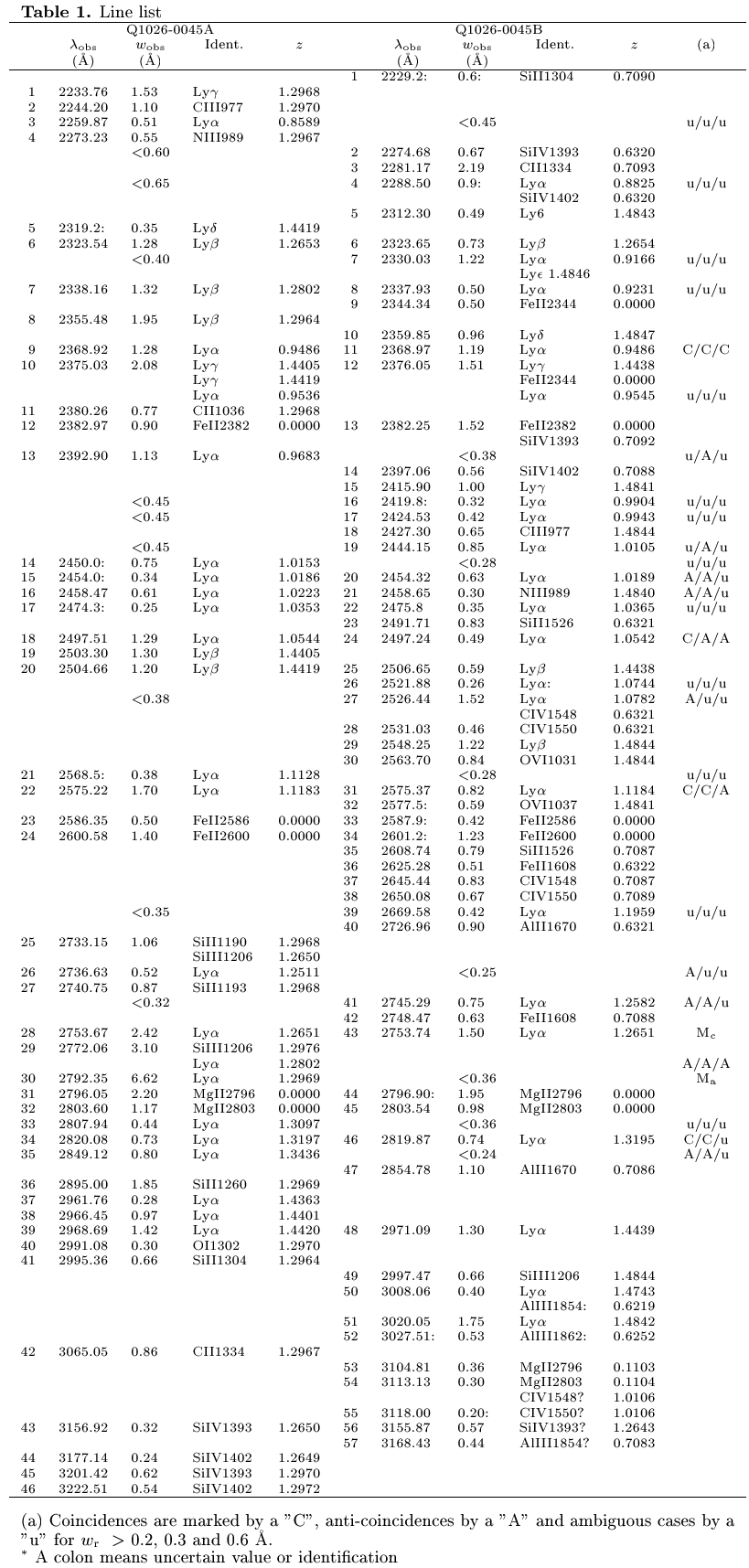,height=23.5cm,width=9.cm,angle=0}}}
\end{figure}

\section{Introduction} \label{intr}
One way to probe the transverse extension of the gaseous structures 
giving rise to the Ly$\alpha$ forest seen in the spectrum of
quasars is to
observe multiple lines of sight to quasars with small angular separations
on the sky and search the spectra for absorptions coincident in redshift.

This technique originated with a suggestion by Oort (1981) to test the
possibility that the Ly$\alpha$ forest clouds originate in large
pancake structures. 
The first discoveries of common and associated 
absorption using pairs of distinct quasars (with separations
$\sim$~1~arcmin) were made by Shaver et al. (1982) and 
Shaver \& Robertson (1983). These already indicated the possible 
existence of very large absorber sizes (hundreds of kpc), even for the 
Ly$\alpha$ clouds. At about the same time Sargent et al. (1982) found 
no detectable tendency for Ly$\alpha$ lines to correlate in QSO
pairs separated by a few arcmin. 
Spectra of pairs of 
gravitational lens images revealed common absorptions on smaller 
scales (Weyman \& Foltz 1983, Foltz et al. 1984).
The idea that Ly$\alpha$ clouds might have large sizes remained controversial
untill the analysis by
Smette et al. (1992), later confirmed by Dinshaw 
et al. (1994), Bechtold et al. (1994), Crotts et al. (1994), Bechtold \& 
Yee (1994), Smette et al. (1995), D'Odorico et al. (1998).

Recently, Dinshaw et al. (1995) derived 
a radius of 330$h^{-1}_{50}$ kpc at $z$~$\sim$~0.7 for spherical clouds 
from observation of
Q0107--0232 and Q0107--0235 separated by 86~arcsec. Larger separations 
have been investigated by Crotts \& Fang (1997) and Williger et al. (1997).
Both studies conclude that the clouds should be correlated on scales
larger than 500~kpc.\par\noindent
Here we present observations of Q1026--005~A 
($m_{\rm r}$~=~18.4, $z_{\rm em}$~=~1.438) and B 
($m_{\rm r}$~=~18.5, $z_{\rm em}$~=~1.520), two distinct quasars
separated on the sky by 36~arcsec or
300~$h^{-1}_{50}$~kpc ($q_{\rm o}$~=~0.5) at $z$~$\sim$~1. 


\section{Observations} \label{s2}
The observations were carried out on the Hubble Space Telescope 
using the Faint Object Spectrograph  with the G270H
grating over the wavelength range 2250--3250~\AA, for a resolution of
1.92~\AA~ FWHM. A total of 5300~s integration time was accumulated
on both quasars.
The data were calibrated using the standard pipeline
reduction techniques. The zero point of the wavelength scale was
determined requiring that Galactic interstellar absorptions occur at
rest. Most of the lines are weak or blended except 
Mg~{\sc ii}$\lambda$2803; the error on the wavelength determination should be 
smaller than 0.3~\AA~ however. 
The spectra are shown in Fig.~1, the line--lists are given in
Table~1. The position of absorption features are determined by gaussian fits.
Lower limits on equivalent widths of Ly$\alpha$ lines are at the
3$\sigma$ level. The mean signal to noise ratio is 15 varying from 10 
in the very blue to 20 on top of the Ly$\alpha$ emission lines.

\section{Results}
\subsection{The metal line systems}
%
Metal line systems are present at $z_{\rm abs}$~=~1.2651 and 1.2969
in A, $z_{\rm abs}$~=~0.11, 0.6320, 0.7090, 1.2651 and 1.4844 in B.\par
In A, the system at $z_{\rm abs}$~=~1.2651 is detected by strong
H~{\sc i}$\lambda\lambda$1025,1215 absorptions and has a
Si~{\sc iv}$\lambda\lambda$1393,1402 doublet associated. There are
strong H~{\sc i}$\lambda\lambda$1025,1215 absorptions at the same
redshift along the line of sight to B. The fact that the 
positions of the H~{\sc i} absorptions
in A and B are nearly identical ($\lambda$2753.67 and
$\lambda$2753.74~\AA~ for Ly$\alpha$ respectively) 
argues for the two absorptions being
produced by the same object. If true the transverse dimension is
larger than the 310~$h^{-1}_{50}$kpc separation between the two 
lines of sight. 
The system at $z_{\rm abs}$~=~1.2969 has strong H~{\sc i} 
($w_{\rm r}$(Ly$\alpha$)~=~2.9~\AA), 
C~{\sc ii}, C~{\sc iii}, O~{\sc i}, N~{\sc iii}, Si~{\sc ii}, Si~{\sc iii},
Si~{\sc iv} absorptions. \par
In B, both $z_{\rm abs}$~=~0.6320 and 0.7090 systems have strong
Si~{\sc ii}$\lambda$1526, Fe~{\sc ii}$\lambda$1608, 
Al~{\sc ii}$\lambda$1670 and C~{\sc iv}$\lambda$1550
absorptions. As said before the $z_{\rm abs}$~=~1.2651 system 
is common to A and B. The presence of metals is revealed only 
by a $\lambda$3155.87 feature that we identify as Si~{\sc iv}$\lambda$1393. 
The associated Si~{\sc iv}$\lambda$ 1402 line is not detected but could be 
lost in the noise. The strong system at $z_{\rm abs}$~=~1.4842 
shows N~{\sc iii}, C~{\sc iii}, Si~{\sc iii} and possibly O~{\sc vi} 
associated absorptions.
O~{\sc vi} absorption seems to be detected in most of the
low and intermediate redshift systems (Bergeron et al. 1994,
Vogel \& Reimers 1995, Burles \& Tytler 1996) and has been observed
in a few high redshift Lyman limit systems
(Kirkman \& Tytler 1997).
Although the velocity difference with the quasar is larger than
4000~km~s$^{-1}$, there is a possiblility that the system is associated 
with the quasar. In addition we tentatively identify a weak 
($w_{\rm r}$~$\sim$~0.3\AA) Mg~{\sc ii} system at $z_{\rm abs}$~=~0.11 in B.
Mg~{\sc ii}$\lambda$2803 is possibly blended with C~{\sc iv}$\lambda$1548 of
a possible weak C~{\sc iv} system at $z_{\rm abs}$~=~1.0106.

\begin{figure}
\centerline{\vbox{
\psfig{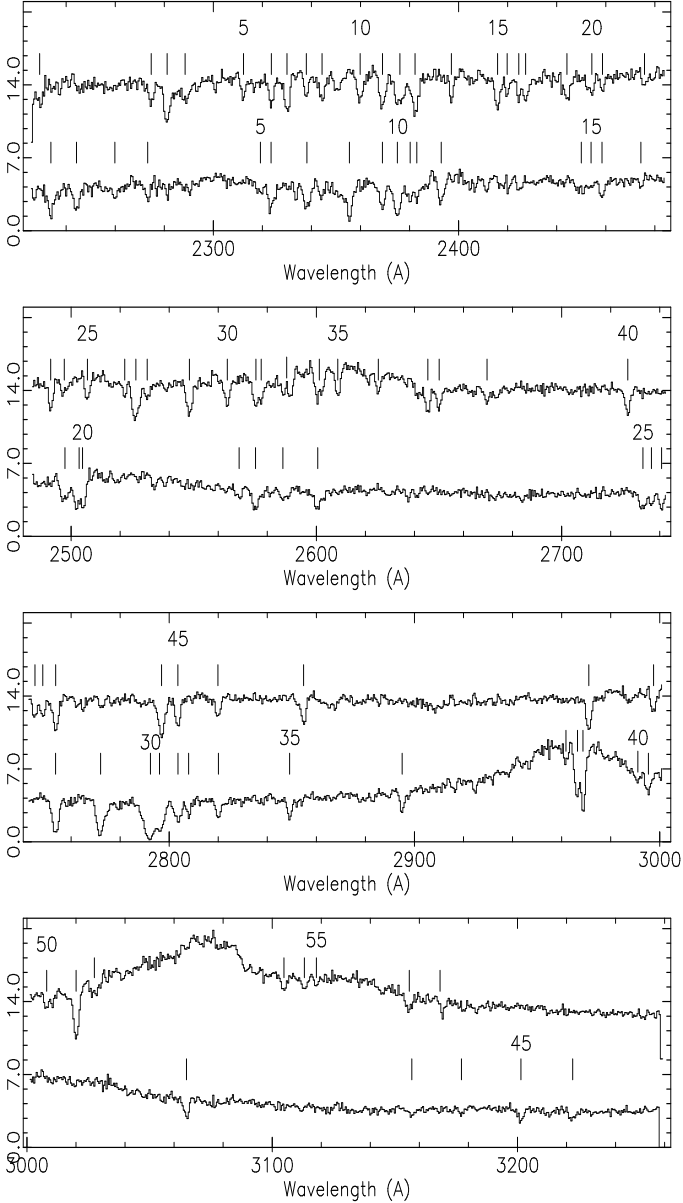}
}}
\caption[]{Spectra of Q1026--0045 A (bottom) and B (top).
Flux is given in units of 10$^{-16}$ erg/s/cm$^{-2}$/\AA.
The spectrum of B has been shifted by
8.5 in the same units. Line identification can be found in Table~1. 
}
\end{figure}

\subsection{The associated system in A and the proximity effect}
There is a strong associated system detected in A by its H~{\sc i}
Ly$\alpha$ and Ly$\beta$ absorptions. Two components are seen 
at $z_{\rm abs}$~=~1.4401 and 1.4420 with 
$w_{\rm r}$(Ly$\alpha$)~=~0.58 and 0.40~\AA~ respectively. 
The two components are 
{\sl redshifted} relative to the QSO by 260 and 490~km~s$^{-1}$. Since the
true redshift of the quasar is poorly known, these values are very uncertain.
We do not detect any metal lines in the system. The O~{\sc vi} lines
have $w_{\rm r}$~$<$~0.20~\AA; the C~{\sc iv} lines are redshifted outside
the wavelength range of the data. Interestingly enough, there is a
H~{\sc i} absorption system at $z_{\rm abs}$~=~1.4439 along the line of sight
to Q~1026--0045B. The velocity difference between this system and
the $z_{\rm abs}$~=~1.4420 in A is about 230~km~s$^{-1}$ only. \par
It is unlikely that the $z_{\rm abs}$~$\sim$~$z_{\rm em}$ 
system is intrinsically associated with the central AGN. Such systems usually
have high metal content and are expected to exhibit strong O~{\sc vi} and 
N~{\sc v} absorptions (Petitjean et al. 1994, Hamann 1997), absent
from the spectra of Q1026--0045~A \& B. The three absorptions
are thus part of an object or group of objects which transversal dimension
exceeds the 300$h^{-1}_{50}$~kpc separation between the two lines of 
sight.\par\noindent
The absence of metals in the system associated with A, over the
observed wavelength range, suggests an
intergalactic origin. The higher velocity of the gas along the line of sight
to B argues against the simple picture in which the gas would be 
collapsing toward A. In that case, we would expect the gas along the 
line of sight to B to have a projected velocity 
smaller than the velocity of the gas just in front of A.
A model where the gas would be part of a rotating disk can be accomodated
if the component at $z$~=~1.4420 is at the same redshift as the quasar. 
In this case however one could wonder why the gas is metal 
deficient.\par\noindent
The relative equivalent widths of the hydrogen lines 
in the Lyman series of the system at $z_{\rm abs}$~=~1.4842 toward B
are indicative of H~{\sc i} column densities in excess of 10$^{16}$~cm$^{-2}$. 
The presence of strong metal lines suggests that the gas is associated with 
the halo of a galaxy. Very deep imaging in this field to search for any 
enhanced density of objects
would help to understand the nature of these intriguing systems.

There are only two systems in both lines of sight
from $z$~=~1.3436 to 1.520, the associated system in A (and 
its counterpart in B) and the metal system at $z$~=~1.4842 in B.
The number of lines with $w_{\rm r}$~$>$~0.24~\AA~
expected in this redshift range is 7$\pm$2 (Bahcall et al. 1996).
It is probable that we see the effect of the enhanced photo-ionizing
field due to the proximity of the quasars. 
%
\subsection{The Ly$\alpha$ forest}
\subsubsection{The line-lists}
Table~1 lists all the absorption features detected at the 3$\sigma$ level 
in the spectra. Identification of Ly$\alpha$ lines is sometimes uncertain
due to blending with lines from the numerous metal line systems.
We discuss here individual lines.\\
In Q~1026--0045A, the $\lambda$2259 feature could be partly Ly$\epsilon$
from the $z_{\rm abs}$~=~1.4420 system but given the strength of
the other lines in the series, the contribution is most certainly negligible.
There is a broad feature centered at 2452~\AA~ that we decompose into
two components at 2450 and 2454~\AA. This feature is uncertain however.
Galactic Fe~{\sc ii}$\lambda$2374 absorption
should not contribute too much to
the $\lambda$2375 feature that is mostly Ly$\gamma$ at $z_{\rm abs}$~=~1.4405
and 1.4420. The line is quite strong however and could be partly produced 
by a Ly$\alpha$ absorption at $z_{\rm abs}$~=~0.9536. 
The two lines at $\lambda$2458 and $\lambda$2474 could correspond to
a Si~{\sc iv}$\lambda$ $\lambda$1393,1402 doublet at $z_{\rm abs}$~=~0.7639.
The corresponding C~{\sc ii}$\lambda$ 1334 line would be blended with 
Ly$\beta$ at $z_{\rm abs}$~=~1.2964 but the feature at $\lambda$2736
could be C~{\sc iv}$\lambda$1550 at the same redshift with no
C~{\sc iv}$\lambda$1548 detected. 
The line however is displaced by more than 1~\AA~ 
from the expected position which is not acceptable. We thus consider
the Si~{\sc iv} identification as doubtful.\\
In Q~1026--0045B, there is a broad feature at $\lambda$2288.5 that
cannot be accounted for by Si~{\sc iv}$\lambda$1402 at $z$~=~0.6320 only.
The feature at $\lambda$2376 may have a double structure. 
Ly$\gamma$ at 1.4438 and Fe~{\sc ii}$\lambda$2374
definitively contribute to this feature which is strong enough
however to be partly produced by
Ly$\alpha$ absorption at $z_{\rm abs}$~=~0.9545.
Since C~{\sc iv}$\lambda$1550 at $z_{\rm abs}$~=~0.6321 has
$w_{\rm obs}$~=~0.46~\AA, 
C~{\sc iv}$\lambda$1548 at the same redshift should have 
$w_{\rm obs}$~$<$~0.92~\AA. Consequently there is
a Ly$\alpha$ line at $z_{\rm abs}$~=~1.0782 with $w_{\rm obs}$~$>$~0.6~\AA.

We detect 11 and 12 Ly$\alpha$ lines with $w_{\rm r}$~$>$~0.2~\AA~ 
over the redshift range 0.8335--1.3436 along the lines of sight to 
A and B respectively. The density of lines with $w_{\rm r}$~$>$~0.24~\AA~
detected by the HST in the same redshift range is $\sim$17$\pm$3
(Jannuzi et al. 1998).
The number of lines we detect is thus small. This might be a consequence 
of blending effects. 
Two lines observed along A and B are said 
coincident when their redshifts are within 200~km~s$^{-1}$.
The Lyman$\alpha$ forest is sparse at low redshift which implies that the
probability for random coincidence is negligible
(only 0.05 for $w_{\rm r}$~$>$~0.2~\AA).
Last column of Table~1 indicates for each Ly$\alpha$ line with
$w_{\rm r}$~$>$~0.2, 0.3, 0.6~\AA~ whether there is coincidence (C) or 
anti-coincidence (A). A letter (u) marks lines that are out of the 
sample or uncertain cases because of blending effets.
%
\subsubsection{Correlations}\label{s3}
The numbers of coincidences and anticoincidences for $w_{\rm r}$~$>$~0.2, 0.3
and 0.6~\AA~ are 4, 3, 1 and 7, 8, 3 respectively. 
Assuming that the Ly$\alpha$ clouds are spheres of radius $R$,
we calculate the probability density for $R$ (see Fig.~2) 
following Fang et al. (1996). The peak of the probability is at 
$R$~=~267, 305 and 364$h^{-1}_{50}$~kpc for $w_{\rm r}$~$>$~0.6, 0.3
and 0.2~\AA. There is a hint for
the dimensions of the structures to be larger for smaller equivalent widths.
This property is expected in simulations (Charlton et al. 1997).
However as shown by Fang et al. (1996) and Crotts \& Fang (1997), 
the radius determined by this method increases with the separation
of the lines of sight indicating that the assumption of a single structure
size is invalid. This has been recognized to be a characteristic of the 
spatial distribution of the Ly$\alpha$ gas in the simulations (Charlton et
al. 1997). It is clear that better statistics in the data are needed to 
have a better understanding of the structures especially to 
discuss the difference between real size of the clouds and correlation
length (Cen \& Simcoe 1997).

It is intriguing to note that the velocity difference $\Delta V$
between lines coincident in redshift along the two lines of sight 
is small. Considering all the pairs, we find 4 and 8 pairs with 
$\Delta V$~$<$~50 and 200~km~s$^{-1}$ respectively. There is no
pair, even along one single line of sight with 
200~$<$~$\Delta V$~$<$~400~km~s$^{-1}$. This has been shown to favor
disk-like structures (Charlton et al. 1995) 
but should be studied in more detail.

\begin{figure}
\centerline{\vbox{
\psfig{figure=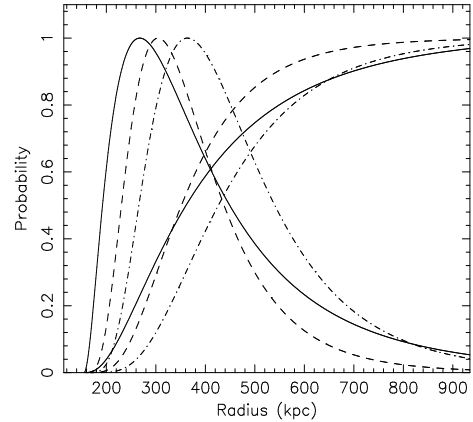,height=7.cm,width=9cm,angle=270}
}}
\caption[]{Probability distribution $P(R)$, normalized 
to one at its peak, and cumulative distribution
versus cloud radius from the number of coincidences and anticoincidences
of Ly$\alpha$ lines with $w_{\rm r}$~$>$~0.2 (dashed-dotted lines), 0.3
(dashed lines) and 0.6~\AA (solid lines). The peak of the probability is at 
$R$~=~267, 305 and 364$h^{-1}_{50}$~kpc for $w_{\rm r}$~$>$~0.6, 0.3
and 0.2~\AA~ respectively.}
\end{figure}


\acknowledgements{MR and JS
wish to thank the SSTC/PRODEX project for partial support during this work.}

\end{document}